\documentclass[aps,prl,twocolumn,groupedaddress]{revtex4}

\usepackage{bm}
\usepackage{graphicx}
\usepackage{color}

\begin{document}

%\preprint{}

\title{Angle-Resolved Photoemission Spectroscopy of the Antiferromagnetic Superconductor Nd$_{1.87}$Ce$_{0.13}$CuO$_4$
: Anisotropic Spin-Correlation Gap, Pseudogap, and the Induced Quasiparticle Mass Enhancement}


\author{
		H. Matsui$^1$,
		K. Terashima$^1$,
		T. Sato$^1$,
		T. Takahashi$^{1}$,
		S.-C. Wang$^2$,
		H.-B. Yang$^2$,
		H. Ding$^2$,
		T. Uefuji$^3$, and
		K. Yamada$^4$}
%\thanks{}
%\affiliation{}
\affiliation{$^1$ Department of Physics, Tohoku University, Sendai 980-8578, Japan}
\affiliation{$^2$ Department of Physics, Boston College, Chestnut Hill, MA 02467, USA}
\affiliation{$^3$ Institute of Chemical Research, Kyoto University, Uji 611-0011, Japan}
\affiliation{$^4$ Institute of Materials Research, Tohoku University, Sendai 980-8577, Japan}


\date{\today}

\begin{abstract}
We performed high-resolution angle-resolved photoemission spectroscopy on Nd$_{1.87}$Ce$_{0.13}$CuO$_4$, 
which is located at the boundary of the antiferromagnetic (AF) and the superconducting phase.  
We observed that the quasiparticle (QP) effective mass around ($\pi$, 0) is strongly enhanced due to the opening 
of the AF gap.  The QP mass and the AF gap are found to be anisotropic, with the largest value near the intersecting point 
of the Fermi surface and the AF zone boundary.  In addition, we observed that the QP peak disappears around the N\a'{e}el 
temperature ($T_N$) while the AF pseudogap is gradually filled up at much higher temperatures, possibly due to the 
short-range AF correlation.
\end{abstract}

\pacs{74.72.Jt, 74.25.Jb, 79.60.Bm}
%\keywords{}

\maketitle
Since the discovery of cuprate high-temperature superconductors (HTSCs), intensive experimental and theoretical 
studies have been performed to elucidate the origin and mechanism of the anomalously high superconducting (SC) 
transition temperature.  It is now widely accepted that electrons or holes doped into the parent Mott insulator 
interact antiferromagnetically with each other on the quasi-two dimensional CuO$_2$ plane.  Although 
it has been suggested that the antiferromagnetic (AF) interaction plays an essential role for pairing of electrons 
(holes) in the SC state, it is still unclear how the antiferromagnetism interplays with the superconductivity at 
the microscopic level.  This problem is a central issue not only in HTSCs but also in other exotic superconductors 
such as heavy-fermion and organic-salt superconductors.  In the phase diagram of electron-doped cuprates, 
the AF and SC phases are adjacent to each other or somewhat overlap at the boundary, in contrast to the hole-doped 
case where the two phases are well separated \cite{Luk}.  The proximity or overlapping between the AF and SC phases in 
the electron-doped cuprates yields a good opportunity for studying the interplay between the AF interaction and the 
superconductivity \cite{Kan,Yam,Fuj,Kuk}.  In fact, a recent elastic neutron-scattering experiment 
reported a competitive nature between the AF long-range order and the superconductivity \cite{Kan}.  On the other 
hand, an inelastic neutron scattering experiment observed coexistence of the gapped commensurate spin fluctuation 
and the superconductivity \cite{Yam}.  In contrast to these intensive studies on the ``{\it q}-resolved" spin 
dynamics by neutron scattering, a limited number of photoemission studies on the ``{\it k}-resolved" electronic structure 
have been reported for electron-doped cuprates \cite{ArL,ArB}.  The {\it k}-dependence of the AF correlation effect on the 
electronic structure is essential to understand the interplay between the antiferromagnetism and the 
superconductivity. 

In this Letter, we report high-resolution angle-resolved photoemission spectroscopy (ARPES) on electron-doped 
cuprate Nd$_{1.87}$Ce$_{0.13}$CuO$_4$ (NCCO, x = 0.13) located at the phase boundary between the AF and SC phases.  We found 
the mass-renormalized quasiparticle (QP) state near ($\pi$, 0), which gradually evolves into the high-energy 
gap \cite{ArL} around the hot spot.  The observed continuous evolution of the electronic structure near the Fermi level ($E_F$) as a function 
of momentum ({\it k}) is explained basically in terms of the band folding caused by the AF ordering.  However, we also found 
a deviation from this simple picture in the {\it k}-dependence of the AF-gap size indicative of non-uniform AF 
scattering in {\it k}-space.  We observed that the QP spectrum shows remarkable temperature dependence in 
accordance with the spin correlation.

High-quality single crystals of NCCO (x = 0.13) were grown by the traveling solvent floating zone method and were 
heat-treated in Ar-gas flow at 900$^{\circ}$C for 10h.  The $\mu$SR and magnetic susceptibility measurements show that the 
crystal is antiferromagnetic below 110 K ($T_N$, N\a'{e}el temperature) and shows a signature of superconductivity below 
20 K ($T_c$) \cite{Uef}.   ARPES measurements were performed with 
GAMMADATA-SCIENTA SES200 and SES2002 spectrometers at Tohoku University and the undulator 4m-NIM beamline 
of Synchrotron Radiation Center, Wisconsin, respectively.  We used monochromatized He I$\alpha$ resonance line 
(21.218 eV) and 22-eV photons to excite photoelectrons.  The energy and angular (momentum) resolutions 
were set at 11 meV and 0.2$^{\circ}$ (0.01\AA$^{-1}$), respectively.  A clean surface of sample for ARPES measurements was 
obtained by {\it in-situ} cleaving along the (001) plane.  The Fermi level of sample was referred to that of a gold film evaporated 
onto the sample substrate.

Figure 1 shows the plot of ARPES intensity near $E_F$ at 30 K as a function of the two-dimensional wave vector to 
illustrate the Fermi surface (FS) of NCCO (x = 0.13).  Bright areas correspond to the experimental FS.  We normalized the intensity 
with respect to the highest binding energy of spectrum (400 meV) \cite{Bori}.
We find in Fig.1 that the ARPES intensity at $E_F$ shows a characteristic {\it k}-dependence while the experimental FS looks circle-like 
centered at ($\pi$, $\pi$) as predicted from the LDA band calculation \cite{Mas}.
On the experimental FS, the strongest ARPES intensity appears near ($\pi$, 0), and a weak but observable intensity is 
seen around ($\pi$/2, $\pi$/2), while there is negligible or no intensity between these two momentum regions.  It is 
noted that the area with negligible ARPES intensity on the FS coincides with the hot spot, namely the 
intersecting point of the LDA-like FS and the AF zone boundary.  
A similar ARPES-intensity modulation has been reported in the case 
of x = 0.15 with different photon energies \cite{ArL}, suggesting that the observed intensity modulation is not due to the 
matrix-element effect.

\begin{figure}
\includegraphics[width=3.4in]{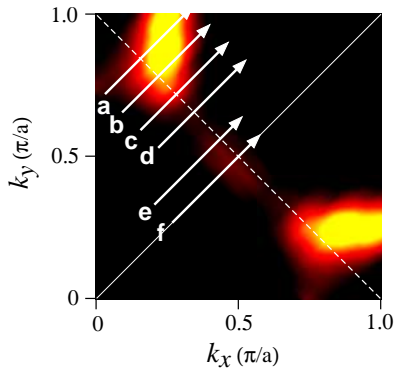}%
\caption{Plot of near-$E_F$ ARPES intensity at 30 K integrated within 25 meV with respect to $E_F$ and symmetrized with respect to the 
(0, 0)-($\pi$, $\pi$) nodal line.  Arrows denoted by {\it a-f} show cuts where detailed ARPES measurements shown in Fig. 2 were done.}
\end{figure}

Figure 2 shows the ARPES spectra near $E_F$ measured along several cuts across the FS (cuts {\it a-f} in Fig. 1) and the 
corresponding band dispersions derived from the spectra.  Near the ($\pi$, 0) point (cuts {\it a-c}), we find two separated 
band dispersions; one is a very steep band dispersion located below $\sim$ 0.1 eV and another is a flat band very close 
to $E_F$. The strong ARPES intensity on the FS near the ($\pi$, 0) point as shown in Fig. 1 is due to this flat band 
located very close to $E_F$.  As seen in Fig. 2, the presence of two separated bands in the 
same momentum region produces the characteristic ``peak-dip-hump (PDH)" structure in the ARPES spectrum 
measured near the Fermi vector ($k_F$).  
The band near 
$E_F$ becomes flatter and the intensity is weakened on going from cut {\it a} to cut {\it c}, namely on approaching the hot spot.  
At the same time, the energy separation between the peak and the hump gradually increases.  In cut {\it d}, which 
passes the hot spot, the peak near $E_F$ almost disappears and as a result an energy gap of about 
100 meV opens between $E_F$ and the lower-lying steep band.  The energy gap 
becomes gradually small on approaching the nodal line and finally the steep band appears to almost touch $E_F$ in 
cut {\it f}, as evidenced by the recovery of the Fermi-edge-like structure in the spectrum.  Here, it is noted that the 
change of band dispersions among different cuts is continuous, indicating that the PDH structure in the ARPES 
spectra in cuts {\it a-c} has a same origin as the energy gap in cuts {\it d} and {\it e}.

\begin{figure*}
\includegraphics[width=5.3in]{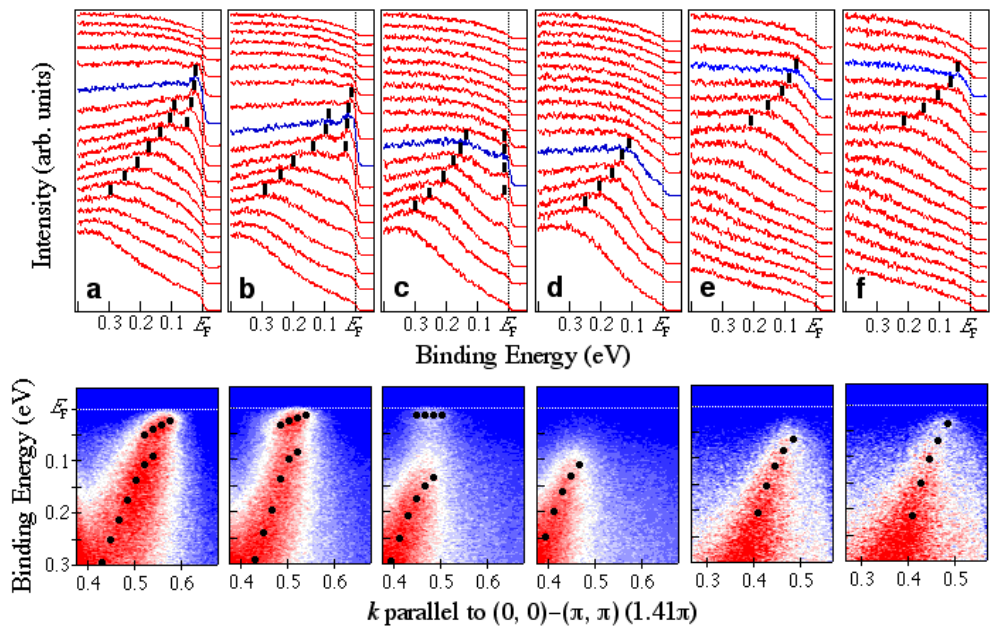}%
\caption{Upper panel: ARPES spectra of NCCO (x = 0.13) measured at 30 K along several cuts parallel to the (0, 0)-($\pi$, $\pi$) 
direction shown by arrows in Fig. 1.  Blue spectra are at the Fermi surface.  Lower panel: ARPES-intensity plot as a function of 
the wave vector and binding energy, showing the experimental band dispersion.  Peak positions in ARPES spectra are shown by 
bars and dots.}
\end{figure*}

The opening of a large energy gap at $E_F$ in cut {\it d} is attributed to the AF spin-correlation, since it is located at 
the intersecting point of the ``original" FS and the ``shadow" FS produced by the AF interaction \cite{ArL}. The present 
ARPES results in Fig. 2 clearly show that the gap at the hot spot is smoothly connected to the two separated bands 
near the ($\pi$, 0) point, suggesting the effect of the AF correlation to modify the band dispersion. We show in Fig. 3 a 
schematic diagram to explain how the QP dispersion is modified by the AF electron correlation.  It is reminded that 
the intersecting point between the original band and the shadow band folded back into the magnetic Brillouin zone 
is always on the diagonal line [($\pi$, 0)-(0, $\pi$)], and more importantly, the intersecting point is {\it below} $E_F$ at 
($\pi$, 0) and {\it above} $E_F$ at ($\pi$/2, $\pi$/2) in the presence of a nearly half-filled circle-like FS centered at 
($\pi$, $\pi$) as shown in Fig. 3.  
The relative position of this intersecting point with respect to $E_F$ plays an essential role in characterizing the 
band dispersion and the ARPES intensity on the FS.  In the case I in Fig. 3, where the intersecting point 
is below $E_F$, the strong AF scattering splits the original dispersion into two pieces above and below the 
intersecting point, respectively, producing an energy gap between the two separated bands.  It is expected that 
the occupied band just below $E_F$ is strongly bent and the QP effective mass is remarkably enhanced in this 
momentum region.  This effect becomes stronger when one approaches the hot spot, because the intersecting 
point with the strongest AF scattering is gradually shifted to $E_F$.  In the case II, where the intersecting 
point is just on $E_F$, the AF scattering eliminates the electronic states at $E_F$, producing a large energy gap 
at $E_F$.  
Finally in the case III, where the intersecting point is above $E_F$, the AF interaction affects the band dispersion 
mainly in the unoccupied states, leaving the original band dispersion in the occupied states almost unaffected. 
We find that the gross feature of band dispersions in different cuts in Fig. 2 shows a good agreement with this 
simple picture.  For example, cuts {\it a}, {\it d}, and {\it f} correspond to the cases I, II, and III, respectively.  Thus 
the observed heavy-mass QP state around ($\pi$, 0) and the {\it k}-dependence of QP dispersion are well explained in 
terms of the effect from the AF correlation.  The continuous evolution of band dispersion along the FS 
shown in Fig. 2 strongly suggests the AF origin of the energy gap and the resulting mass-enhancement in NCCO.

We find in Fig. 2 that the energy separation between the peak and 
the hump (namely the energy separation between the upper and the lower bands separated by the AF correlation) 
gradually increases from cut {\it a} ($\sim$ 50 meV) to cut {\it c} ($\sim$ 120 meV).  This {\it k}-dependence of the AF gap is not 
necessarily obvious in the simple picture in Fig. 3 and may suggest that the strength of the AF scattering has momentum 
dependence, with the stronger amplitude close to the hot spot.
A recent tunneling spectroscopy reported a pseudogap comparable in the size to the superconducting gap, suggesting the second order 
parameter hidden within the supercondcting state in electron-doped HTSCs \cite{Alf}.  However, the one-order smaller energy scale 
compared to the AF gap suggests the different nature between these two gaps.

The mass-enhancement effect and the PDH structure in Fig. 2 look similar 
to those in hole-doped HTSCs, which have been interpreted with some corrective modes such as
the magnetic-resonance mode \cite{PDH}.  However, such arguments is not applicable 
to the electron-doped case, because the QP state is clearly observed even above $T_c$.  As described above, 
the mass-enhancement in NCCO is due to the band folding caused by the AF order/fluctuation.  In this case, the energy 
separation between the {\it peak} and the {\it dip} does not reflect the energy of collective mode, but the separation between the 
{\it peak} and the {\it hump} is related to the AF exchange interaction.  
It is also remarked here that the {\it k}-region where the heavy-mass QP state is observed coincides with the {\it k}-region 
where a large {\it d}-wave superconducting gap opens \cite{Gap}.  This suggests that the superconductivity in 
electron-doped HTSCs occurs in the antifferomagnetically correlated QP state \cite{Zhe}.  The present experimental 
result that the QP effective mass at $E_F$ and the AF gap increase as moving away from ($\pi$, 0) suggests a slight 
deviation in the SC order parameter from the simple $d_{x^2-y^2}$ symmetry, $\Delta(k)$$\propto$ cos($k_xa$)-cos($k_ya$), 
in NCCO \cite{Blu,Hir}.

\begin{figure}
\includegraphics[width=3.4in]{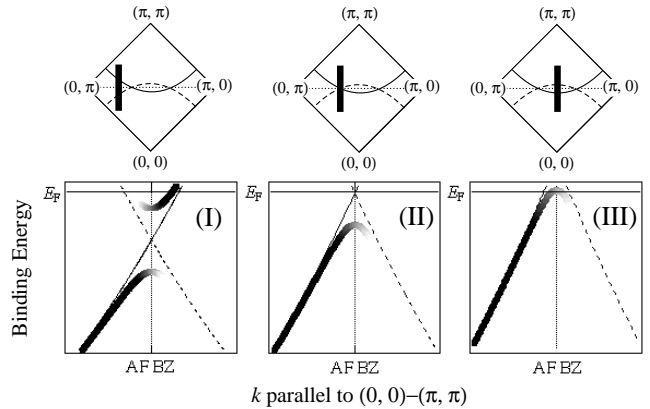}%
\caption{Schematic diagram to explain how the QP dispersion is modified by the AF correlation for three different 
cases.  In case I, the original QP band (thin solid line) and the shadow band (thin broken line) intersect each other below 
$E_F$.  
In cases II and III, the intersecting point is at and above $E_F$, respectively.  Thick solid lines show the QP dispersions 
modified by the AF correlation.}
\end{figure}

\begin{figure}
\includegraphics[width=3.4in]{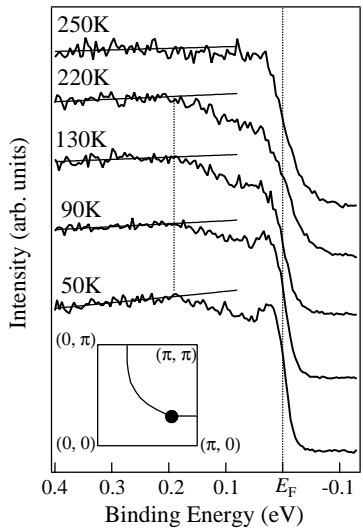}%
\caption{Temperature dependence of ARPES spectrum of NCCO (x = 0.13) measured at a point on the Fermi surface shown by a 
filled circle in the inset, where the PDH structure is clearly observed.  Solid straight lines on the spectra show the linear 
fits to the high-energy region (0.2 - 0.5 eV).}
\end{figure}

Next, we discuss how the heavy-mass QP state at cuts {\it a-c} in Fig. 2 changes as a function of temperature.  Figure 4 
shows the temperature dependence of ARPES spectrum measured at a point on the FS where the PDH structure is 
observed (see the inset to Fig. 4).  At low temperatures (50 and 90 K) below $T_N$ (110K), we clearly find a relatively 
sharp QP peak at $E_F$ and a broad hump at 190 meV in the ARPES spectra, which are ascribed to the upper and lower 
pieces of the renormalized QP band, respectively.  On increasing the temperature across $T_N$, the QP peak 
at $E_F$ becomes substantially broadened and almost disappears at 130K. However, the suppression of the spectral 
weight near $E_F$ ($E_F$ - 0.2 eV), which defines ``a large pseudogap", is still seen in the spectra at 130 K and 220 K. 
This suggests that while the long-range spin ordering disappears at $T_N$, the short-range spin-correlation survives 
even above $T_N$, affecting the electronic structure near $E_F$.  It is remarked that the energy of hump (190 meV) does 
not change with temperature.  On further increasing the temperature, the pseudogap is totally filled-in in the 
spectrum at 250 K, suggesting that the short range AF correlation disappears at around this temperature.  The optical 
conductivity experiment has reported that a pseudogap-like suppression starts to develop in the energy-range 
lower than 0.18 eV at 190 K for x = 0.125 \cite{Ono}, consistent with the present study.  
Further, the optical conductivity and the 
Raman experiments have reported the simultaneous evolution of both the low-energy Drude-like response and 
the high-energy gap on decreasing the temperature \cite{Ono,Koi}.  This shows a good correspondence to the gradual 
development of the QP peak at $E_F$ in ARPES spectrum at low temperatures. The reported sharpening of the low 
energy optical response in NCCO \cite{Ono,Koi} is well explained in terms of the QP mass-enhancement due to the AF 
correlation.

In conclusion, we have performed high-resolution ARPES study on electron-doped cuprate Nd$_{1.87}$Ce$_{0.13}$CuO$_4$.  We 
observed a systematic variation of the QP band dispersion as a function of momentum in addition to the 
characteristic ARPES-intensity modulation on the FS.  These experimental results are well explained in 
terms of the {\it k}-dependent band-folding effect due to the AF ordering.  We observed that the effective QP mass 
around ($\pi$, 0) is strongly renormalized due to the opening of the AF gap.  The QP mass and the AF gap are anisotropic, 
with the largest value near the hot spot, and with a smaller 
value in the vicinity of ($\pi$, 0), suggesting the stronger AF scattering at the hot spot than at ($\pi$, 0).  
Temperature-dependent measurements show that the QP peak gradually weakens on increasing temperature and 
disappears at around the N\a'{e}el temperature, while the AF pseudogap defined by the hump structure is seen 
at temperatures much higher than $T_N$, suggesting that the short-range AF correlation survives well above $T_N$.

We thank H. Ikeda and Kosaku Yamada for valuable discussion. This work was supported by grants from the MEXT of 
Japan, US NSF DMR-0072205, and  US petroleum Research Fund.  H.M. thanks a financial support from JSPS.  The Synchrotron Radiation Center is 
supported by US NSF DMR-0084402.
%\end{acknowledgments}

\newpage


\end{document}